\documentclass[prl,aps,twocolumn]{revtex4}
\usepackage{psfig}
\usepackage{mathrsfs}

\begin{document}
\title{Feigenbaum Cascade of Discrete Breathers in a Model of DNA}
\author{P. Maniadis}
\author{B. S. Alexandrov}
\author{A. R. Bishop}
\author{K. \O. Rasmussen}
\affiliation{Theoretical Division,
Los Alamos National Laboratory, Los Alamos, New Mexico 87545.}
\date{\today}
\begin{abstract}
We demonstrate that period-doubled discrete breathers appear from the anti-continuum limit of the driven Peyrard-Bishop-Dauxois model of DNA. These novel breathers result from a stability overlap between sub-harmonic solutions of the driven Morse oscillator. Sub-harmonic breathers exist whenever a stability overlap is present within the Feigenbaum cascade to chaos and therefore an entire cascade of such breathers exists. This phenomenon is present in any driven lattice where the on-site potential admits sub-harmonic solutions.  In DNA these breathers may have ramifications for cellular gene expression.
\end{abstract}
\maketitle
{\em Introduction.---} Discrete breathers are spatially localized, temporally periodic
excitations in nonlinear lattices \cite{FG}. While discrete breathers share many traits 
with solitons they stand out because of their localization brought about by a delicate sensitivity to 
lattice discreteness. Discrete breathers, have been ubiquitously studied in a wide variety of physical
systems and have been the subject of intense theoretical and numerical scrutiny \cite{FG,Aubry,GST, MJ1,MJ2}.
Although, dissipation and driving are typically key experimental features, most of the theoretical and numerical 
studies have excluded such effects. Only quite recently have the effects of dissipation and driving been 
explicitly considered in numerical studies of discrete breathers \cite{PM1}.

Biomolecules represent a striking example of systems where the localization of discrete breathers 
has long been considered important, and where dissipation and driving cannot be ignored. DNA represents a specific biomolecular system where 
localization of energy in terms of strand-separation dynamics is emerging as a  governing regulatory factor \cite{Regulatory}. The Peyrard-Bishop-Dauxois (PBD) 
model of double-stranded DNA \cite{2,2.1} is arguably the most successful model for describing this local pairing/unpairing (breathing) dynamics, since it reproduces a wide variety of experiments
related to strand-separation dynamics \cite{6}. Of particular importance is the role this model has played in demonstrating strong correlations between regulatory activity, such as protein binding and
transcription, and the equilibrium propensity of double-stranded DNA for local strand separation \cite{-8,3,4,5}.  Recently, this model 
was augmented to include a monochromatic drive in the terahertz frequency range, and was suggested to represent a simplified model for DNA dynamics in the presence of terahertz radiation \cite{PLA}. In this recent work \cite{PLA} the existence 
of discrete breathers oscillating at half the frequency of the drive was numerically observed, and it was argued that the existence and spontaneous generation of such breathing states 
may have significant ramifications for cellular gene expression: This argument has received some experimental support \cite{Berkeley, THZ}. 

Here, we investigate these period-doubled breather excitations further, and 
show that they appear naturally from the anti-continuum limit \cite{MA} in systems of nonlinear oscillators where multi-stability occurs between various stages of the Feigenbaum 
period doubling cascade. Although, our study focuses on the PDB model the observed phenomena are rather general and not limited to this specific system of nonlinear oscillators.

{\em DNA model.---}We consider the PBD model in the following form \cite{PLA}:
\begin{eqnarray} \label{eqn2}
m \ddot y_n = &-& U^\prime (y_n) - W^\prime (y_{n+1},y_{n})
- W^\prime (y_{n},y_{n-1})\nonumber \\ &- &m\gamma \dot y_n +F_0\cos{\Omega t},
\end{eqnarray}
where the Morse potential 
\begin{eqnarray} 
U (y_n)=D \left [\exp {(-ay_n)} -1 \right ]^2 \nonumber
\end{eqnarray}  
represents the hydrogen bonding of the complementary bases. 
Similarly, 
\begin{eqnarray} 
W (y_n, y_{n-1})=\frac{k}{2}\left (1+\rho e^{-\beta(y_n + y_{n-1}) } \right ) \left (y_n-y_{n-1} \right )^2  \nonumber
\end{eqnarray}
represents the stacking energy between consecutive base pairs.  
The parameters $D$, $a$, and $k$ in principle depend on the type of the base pair (A-T or G-C), but for simplicity we will consider homogeneous DNA in this study.
The term $m\gamma \dot y_n$ is the drag caused by the solvent, while $F_0\cos{\Omega t}$ is the (terahertz) drive \cite{note_on_parameters}. In this system, Eq. (\ref{eqn2}),
the linear resonance frequency is given as $\Omega_0^2=2Da^2/m$.
\begin{figure}[h]
\psfig{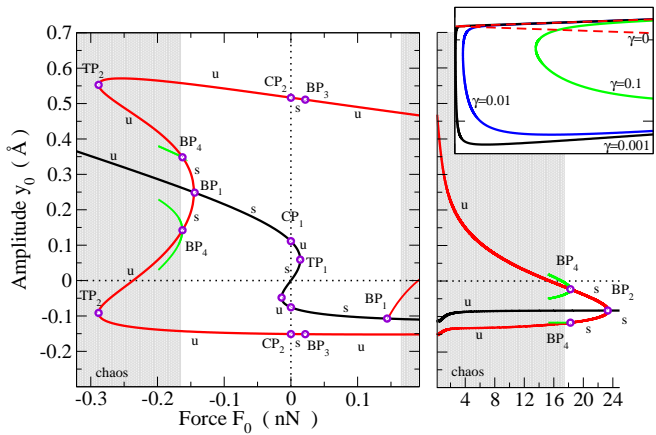}
\caption{(Color online) Nonlinear response manifold of a single oscillator driven at $\Omega=$7.0 rad/ps ($< \Omega_0$). The NLRM for the period-$T$ solutions (solid black line) and the period-doubled solutions (dashed red line) are
shown together with parts of the NLRM for the period-$4T$ (solid blue line segments connected to $BP_4$) solutions. The stability of the various solutions is indicated by the letters placed in the vicinity of the lines ('s' for stable and 'u' for unstable).
Important turning points ($TP$), bifurcation points ($BP$), and crossing points ($CP$) are shown by circles. Finally, the grey areas represent the chaotic regions that result from the 
Feigenbaum period-doubling cascades. The inset shows the behavior near $CP_1$ in the presence of dissipation. $\gamma=0$ (dashed red line), $\gamma=0.001$ps$^{-1}$, $\gamma=0.01$ps$^{-1}$, and $\gamma=0.1$ps$^{-1}$}
\label{FIG1}
\end{figure}

In order to understand how period-doubled breathers oscillating at half the frequency, $\Omega/2$, of the drive can appear at the anti-continuum limit, we first study the nonlinear response of a single ($k=0$) oscillator. For periodic 
driving the single oscillator supports periodic solutions with the period ($T=2\pi/\Omega$) of the drive, and, due to the softness of the Morse potential, this oscillator also supports solutions with 
period $nT$, where $n$ is any positive integer. These solutions are most simply studied at zero dissipation ($\gamma=0$); we will then examine the modifications resulting when $\gamma \neq 0$. A Newton method \cite{Aubry} was used to
track the solutions with periods $T$ and $2T$. The Newton procedure was initiated for a very small driving amplitude ($F_0 \simeq 0$). The response of the system, namely the amplitude of the period $T$ and $2T$ solutions, was
followed as $|F_0|$ was increased. 

{\em Soft period-doubled breathers: $\Omega < \Omega_0$. ---} In the above fashion we constructed the nonlinear response manifolds (NLRM) \cite{KA} depicted for $\Omega=  7$ rad/ps in Fig. \ref{FIG1}. The NLRM for the oscillations with period $T$ is
shown by the solid black line, while the dashed red line represents the NLRM for the period-doubled ($2T$) solutions. Finally, the blue lines segments (connected to $BP_4$) represent parts of the NLRM for the period $4T$ solutions. The stability of the solutions 
can be obtained by Floquet stability analysis \cite{Aubry} and is indicated 
by the letters 's' (stable) and 'u' (unstable) placed adjacent to the respective NLRMs. The stability of a given solution changes at the turning points ($TP$), bifurcation points ($BP$), or at the crossing points ($CP$) of the NLRM.
Since, the Morse potential is asymmetric with respect to $y=0$, the NLRMs lack symmetry with respect to a $\pi$-phase shift ($\pm F_0$) of the drive. However, an equivalence between  the NLRM branches for $F_0$ and $-F_0$ remains, as is 
indicated by our labeling of the TPs, BPs, and CPs.  
\begin{figure}[h]
\psfig{figure=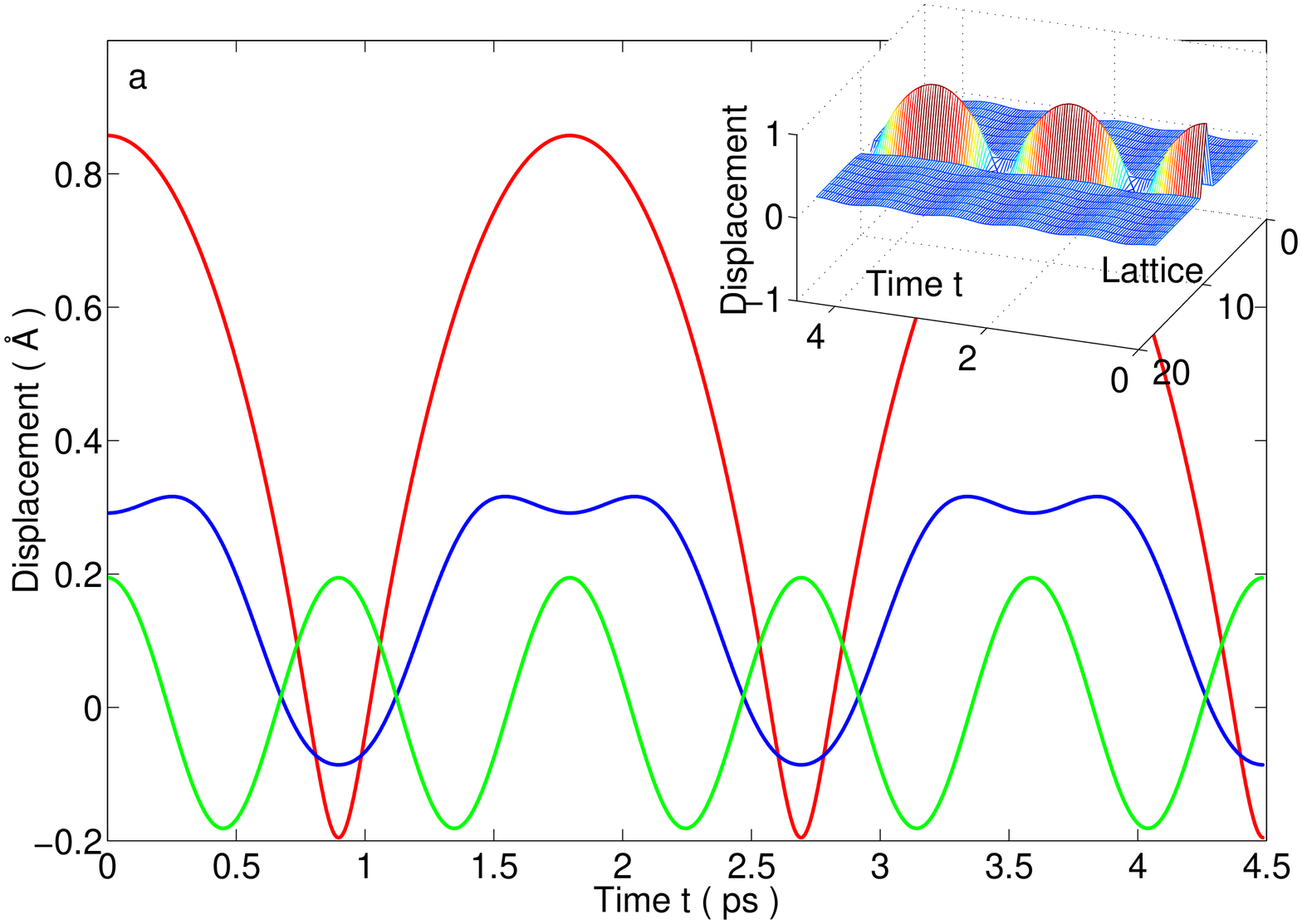,width=\columnwidth,angle=0}
\psfig{figure=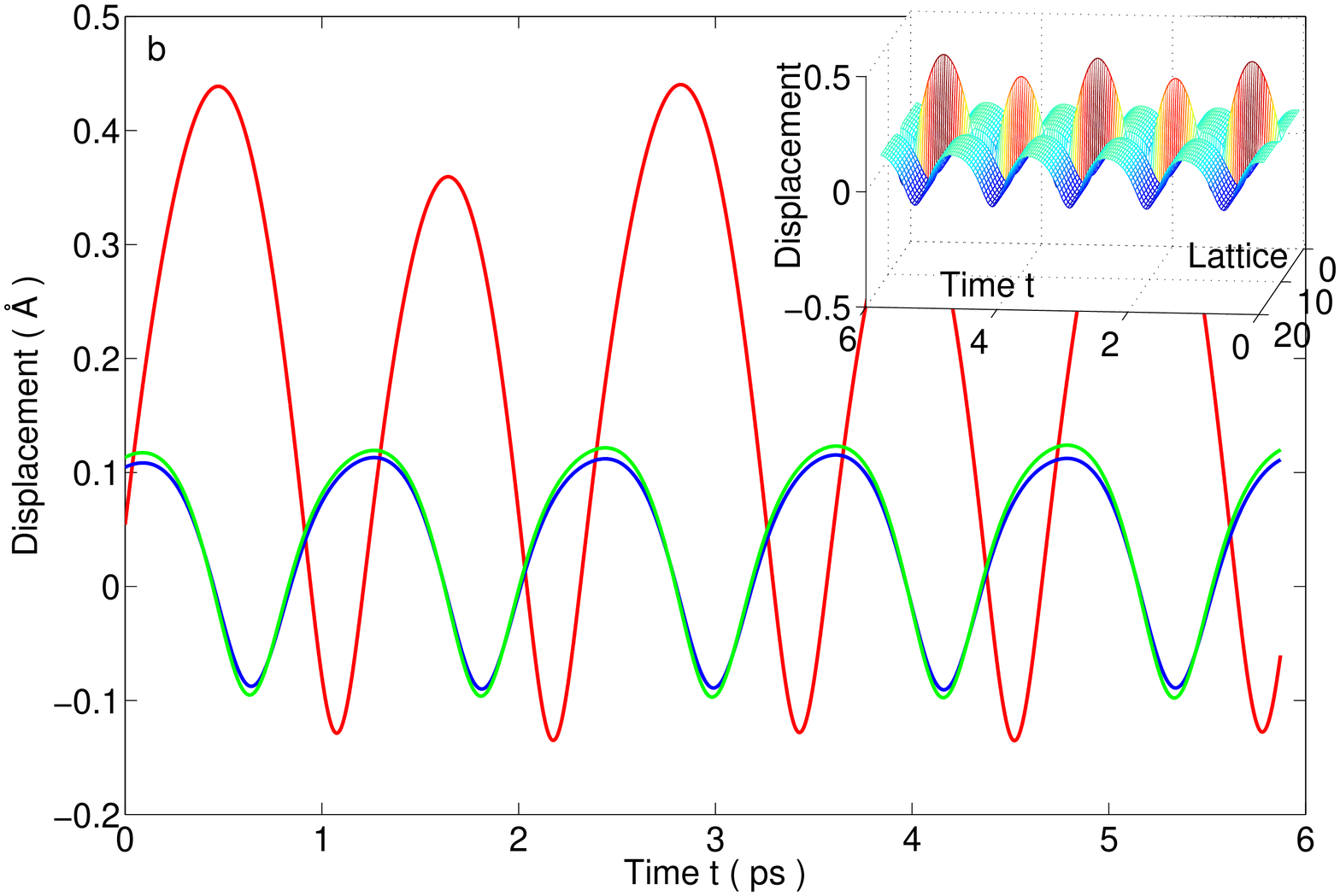,width=\columnwidth,angle=0}
\caption{Illustration of period-doubled breathers. Shown are the central site $y_0$ of the breather (solid red lines) together with the closest neighbor sites $y_1$ (dashed blue lines), and $y_2$ (dashed-dotted green lines). For clarity, panel a)
shows $10y_1$ and $10y_2$, respectively. Panel a) shows the period-doubled breather corresponding to the stability overlap shown in Fig. \ref{FIG1} ($\Omega=7$ rad/ps and $\gamma = 0$) for $F_0=7.06$ pN and $k=0.02$eV$\AA^{-2}$. Similarly,
panel b) shows a period-doubled breather for $\Omega = 5.35$ rad/ps, $\gamma=1$ps$^{-1}$, $F_0 = 115.6$ pN, and $k=0.0065$eV$\AA^{-2}$. The insets shows the full spatio-temporal evolution of the period-doubled breathers. }
\label{FIG2}
\end{figure}

From Fig. \ref{FIG1} it can be noticed that, for a region of small positive force amplitudes, $F_0$, the period-$T$ solution is stable up to $TP_1$ ($F_0 \simeq \pm 14$ pN) and so is the period-doubled solution 
between $CP_2$ ($F_0=0$) and $BP_3$ ($F_0 \simeq 21.5$ pN). This means that in the region between $0 < F_0 < 14$ pN both solutions are stable. A similar stability overlap can be observed for the 
regions $F_0=0$ to $TP_1$ (period-T, $F_0<0$) and $BP_4$ to $BP_1$ (period-doubled) when $\Omega <  5.5$ rad/ps (not shown in Fig. \ref{FIG1}). The ramifications of these stability overlaps for the construction of discrete 
breathers is now clear.  In the anti-continuum limit ($k=0$) of $N$ driven oscillators, we can construct solutions where $N-1$ of the oscillators perform stable period-$T$ oscillations, while the remaining oscillator
resides stably in the period-doubled state. Performing standard numerical continuation \cite{Marin} of this state to finite values of the coupling, $k$, will retain the stability of the solution and 
lead to a stable period-doubled discrete breather solution. 
Examples of such period-doubled breathers 
arising in each of the two described stability overlap regions are shown in Fig. \ref{FIG2}.
Panel a) of Fig. \ref{FIG2} shows the period-doubled breather in the stability overlap region for $F_0 >0$. It is clearly seen that the breather, which extends over $~5$ lattice sites, 
performs oscillations at exactly half the frequency of the background. The background is frequency-locked to the drive at frequency $\Omega$. Similarly, panel b) of  Fig. \ref{FIG2} shows the 
period doubled breather that exists in the stability overlap region for $F_0<0$. The breather is more localized and its dynamical behavior is more complex. This breather roughly follows the background 
period-T oscillations and the period-doubling occurs as a modulation of the breathing amplitude. For weak driving the dynamics must be very similar to the undriven case, as is seem in panel a). At 
stronger driving the dynamics is more dominated by the drive and become more intricate, as in the case of panel b).

It should be noted that, in the absence of a stability overlap, unstable period-doubled breathers can of course still be constructed. An interesting example of this occurs for $F_0 < 23.3$ nN where 
Fig. \ref{FIG1} shows that the period-doubled solution is stable while the period-T solution is unstable. Close to $BP_2$ the period-T solution is very weakly unstable 
(the Floquet exponent is $\sim 1.04$), so that a weakly unstable 'dark' breather can be constructed in this region by positioning N-1 of the oscillators in the period-doubled states while the 
remaining oscillator is in the period-T state. We have found that numerical continuation of this state produces a weakly unstable 'dark' breather that can be sustained for tens of periods.
\begin{figure}[t]
\psfig{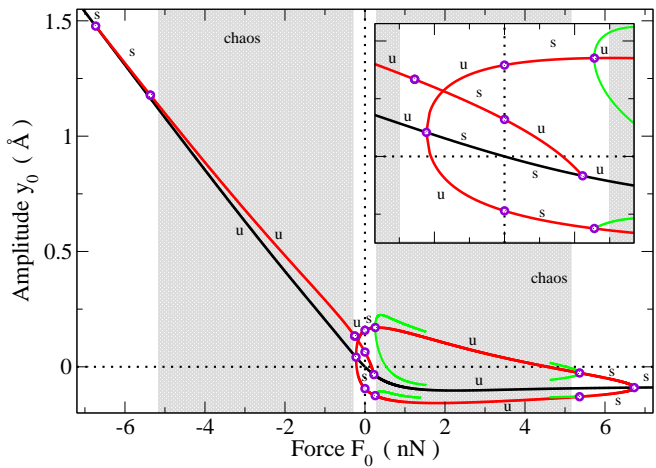}
\caption{Nonlinear response manifold of a single oscillator driven at $\Omega=$13.194 rad/ps ($> \Omega_0$). The NLRM for the period-$T$ solution (solid black line) and the period-doubled solutions (dashedred line) are
shown together with parts of the NLRM for the period-$4T$ (blue line segments) solutions. The stability of the solutions is indicated by the letters placed in the vicinity of the lines ('s' for stable and 'u' for unstable).
Important turning points, bifurcation points, and crossing points are shown by circles. Finally, the grey areas represent the chaotic regions that result from the 
Feigenbaum period-doubling cascades. A stability overlap can be observed in the vicinity of $F_0=0$. The inset shows a magnified version of the area between the 
chaotic regions.}
\label{FIG3}
\end{figure}

The scenario described above generally holds also when dissipation is taken into account. The modifications to the NLRM caused by dissipation is thoroughly discussed for a 
different system of nonlinear equations in Ref. \cite{PM1}. The main effect of dissipation occurs at the crossing points $CP_1$ and $CP_2$. For $\gamma=0$ the period-T solutions are in phase with the 
drive when the response amplitude $y_0$ is below $CP_1$, but it is in anti-phase ($\pi$-phase shifted) when $y_0$ is above $CP_1$. Similarly, the period-doubled solutions experience a $\pi$-phase shift 
at $CP_2$. The introduction of dissipation creates an additional phase shift between the drive and the response everywhere along the NLRM. This additional phase shift is largest close to 
the crossing points and cause the solution to disappear close to the crossing points (small $|F_0|$). This effect is illustrated in the inset of Fig. \ref{FIG1} for the $CP_2$ 
crossing point. It is clear that for weak dissipation the described stability scenario persists. However, as the dissipation increases the stability overlap region is eventually lost and 
with it the stable period-doubled breathers.
\begin{figure}[t]
\psfig{figure=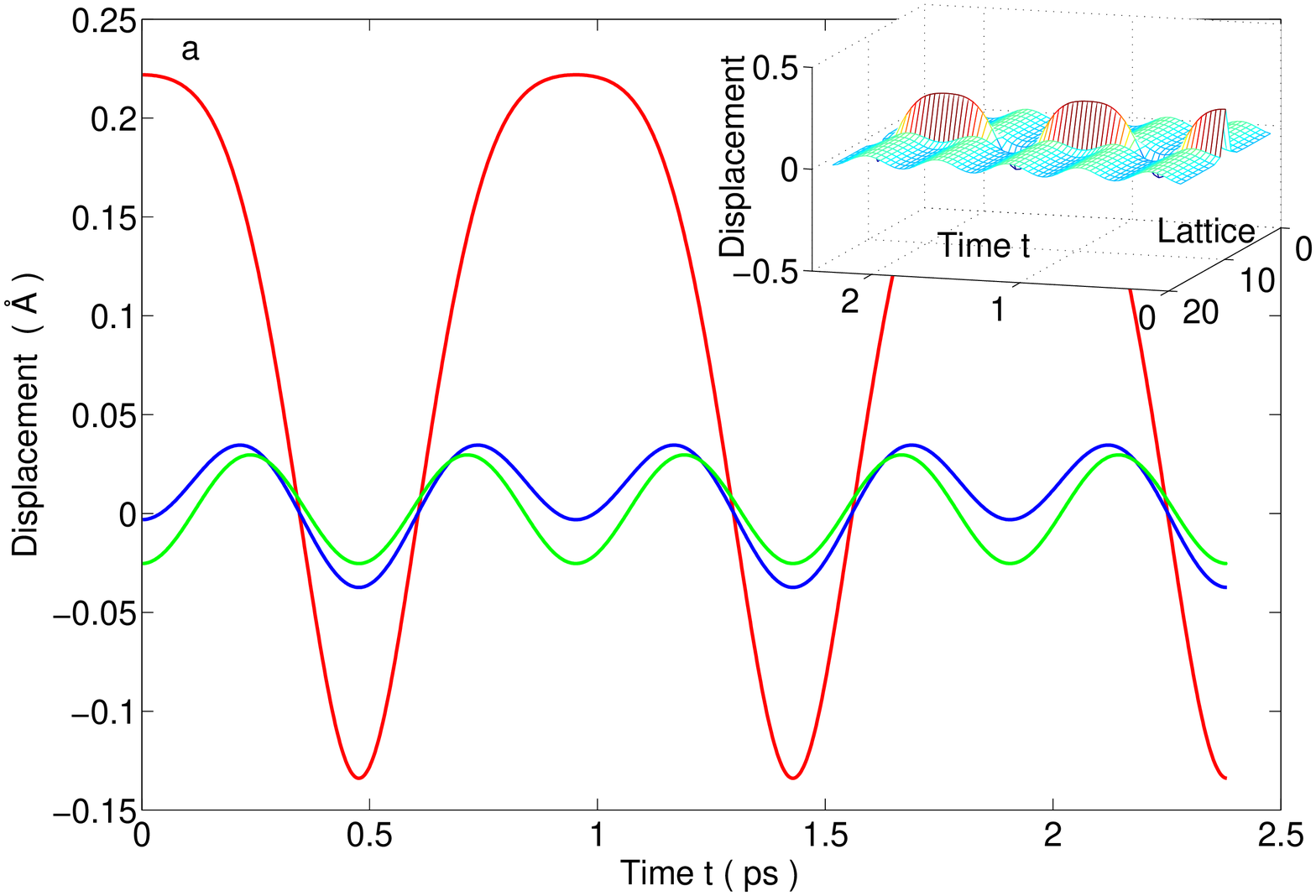,width=\columnwidth,angle=-0}
\psfig{figure=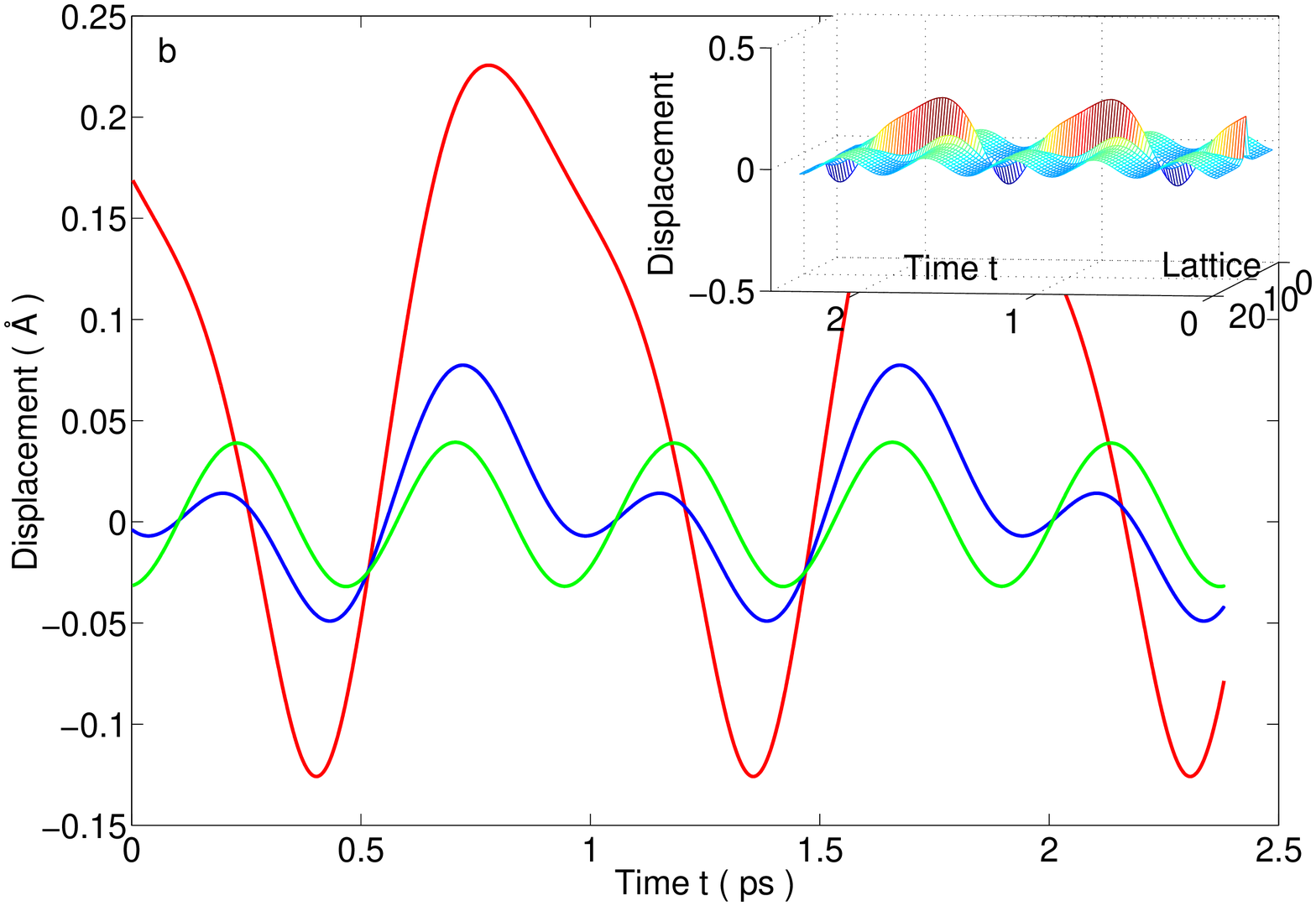,width=\columnwidth,angle=-0}
\caption{Illustration of period-doubled breathers. Shown are the central site $y_0$ of the breather (solid red lines) together with the closest neighbor sites $y_1$ (dashed blue lines), and $y_2$ (dashed-dotted green lines). 
Panel a) shows the period-doubled breather corresponding to the stability overlap shown in Fig. \ref{FIG1} ($\Omega=13.194$ rad/ps and $\gamma = 0$) for $F_0=160$ pN and $k=0.025$eV$\AA^{-2}$. Panel b) 
demonstrates the effects of dissipation $\Omega=13.194$ rad/ps,  $\gamma = 1$ps$^{-1}$, $F_0=209$ pN, and $k=0.025$eV$\AA^{-2}$. The insets show the full spatio-temporal evolution of the period-doubled breathers.} 
\label{FIG4}
\end{figure}

{\em Hard period-doubled breathers: $\Omega > \Omega_0$. ---} Figure \ref{FIG3} shows the NLRM when the frequency of the drive ($\Omega = 13.194$ rad/ps) is larger than the linear resonance frequency of the on-site potential. In this case only a single, but larger, stability overlap between the period-T solution and the period-doubled solution exists in the vicinity of $F_0=0$. Again, the period-doubled breathers can be generated in this overlap region. It is noteworthy
that for soft potentials, such as the Morse potential, ordinary period-T breathers cannot exists above the linear $\Omega_0$ resonance frequency: This frequency region can 
only be accessed for hard potentials. 

Two examples of the period-doubled breathers for $\Omega = 13.194$ rad/ps are given in Fig. \ref{FIG4}. Panel a) shows a period doubled-breather in a dissipation-less ($\gamma=0$) system, and it is clear that the 
breather sites oscillate at half the frequency of the background, which is frequency locked to the drive. Similarly, panel b) shows a period-doubled breather at strong dissipation $\gamma=1$ps$^{-1}$. The phase shift 
introduced by the dissipation is apparent.

{\em Discussion and conclusions. ---} We have demonstrated the existence of a new kind of sub-harmonic discrete breathers in driven lattices of coupled nonlinear oscillators. The phenomenon was illustrated within the 
driven PBD model, which consists of a lattice of coupled Morse oscillators. Specifically, we have shown that these breathers appear naturally from the anti-continuum limit in a 
fashion similar to familiar discrete breathers. We show that the sub-harmonic breathers arise from a stability overlap between sub-harmonic solutions of the single oscillator.
This new kind of breathers exist even when the frequency of the drive is above the linear resonance frequency of the soft on-site potential. Ordinary breather solutions 
do not exist in such cases for soft potentials. 

Although, we focused here on the stability overlap between the frequency locked solution and the period-doubled solution, it is 
clear that in the presence of a stability overlap the phenomena can occur between any
of the sub-harmonic solutions in the Feigenbaum cascade. We have explicitly verified this by constructing stable periodic-$nT$ ($n=3,4,5$, and $6$ ) breathers for small $F_0$ and $k$, in the case
where $\Omega > \Omega_0$. In this fashion a cascade of sub-harmonic discrete breathers exists. The phenomenon is not specific to the system we have studied: it is present in 
any driven lattice of nonlinear oscillators where the on-site potential admits sub-harmonic solutions. The sub-harmonic breathers also exist in the presence of dissipation, it is however 
important to note that in this case the cascade is truncated. 

The existence of the sub-harmonic breathers in DNA is potentially very important because this suggests 
that electromagnetic terahertz radiation can lead to the spontaneous generation of stable local strand separation in double-stranded DNA, and thereby could lead to modification in cellular 
gene expression by affecting transcription and other processes. Such effects have recently been experimentally observed \cite{Berkeley,THZ}. Future extensions of our approach will include effects of
gene sequence and other disorder, as well as temperature.

\acknowledgments This research was carried out under the auspices of the 
National Nuclear Security Administration of the U.S. Department of 
Energy at Los Alamos National Laboratory under Contract No. DE-AC52-06NA25396.

\end{document}